\documentclass[prl]{revtex4}
\usepackage{graphicx}

\begin{document}
\title{Insights into the fracture mechanisms and strength of amorphous 
and nanocomposite carbon}

\author{M.G. Fyta, I.N. Remediakis, and P.C. Kelires}
\affiliation{Physics Department, University of Crete, P.O. Box 2208, 
710 03, Heraclion, Crete, Greece}

\author{D.A. Papaconstantopoulos}
\affiliation{Department of Computational and Data Sciences, George Mason
University, Fairfax, VA 22030, USA}


\begin{abstract}

Tight-binding molecular dynamics simulations shed light into the
fracture mechanisms and the ideal strength of tetrahedral amorphous 
carbon and of nanocomposite carbon containing diamond crystallites, 
two of the hardest materials. It is found that fracture in the 
nanocomposites, under tensile or shear load, occurs inter-grain and so 
their ideal strength is similar to the pure amorphous phase. 
The onset of fracture takes place at weakly bonded $sp^3$ sites in the 
amorphous matrix. On the other hand, the nanodiamond inclusions 
significantly enhance the elastic moduli, which approach those of diamond.

\end{abstract}

\maketitle

The quest for new superhard materials, with hardness and thermal 
stability comparable to diamond, has generated intense research 
at both the fundamental and practical level. Carbon-based materials
are considered to be the best candidates for extreme hardness. 
One of the most widely studied pure carbon materials is tetrahedral amorphous 
carbon ($ta$-C), containing a high fraction of $sp^3$ hybrids, with
many desirable diamondlike properties \cite{Reviews}. The nanohardness
of $ta$-C with 85-90\% $sp^3$ bonding has been measured to range from 
$\sim$ 50-60 GPa \cite{Pharr-Shi-Martinez} to $\sim$ 80 GPa \cite{Friedmann}, 
compared to 100 GPa for diamond, which makes it one of the hardest 
materials. 

Nanocomposite carbon is expected to further improve the mechanical 
properties of pure amorphous phases through the appropriate choice of 
nanoinclusions. Such a prototypical nanocomposite carbon material consists 
of diamond crystallites embedded in a $ta$-C matrix. The stability of this 
phase, denoted by $n$-D/$a$-C, was theoretically predicted by 
Fyta {\it et al.} \cite{Fyta03}. Its synthesis and growth mechanisms have 
been explored by Lifshitz and co-workers, in both hydrogenated \cite{Lifshitz}
and pure \cite{Yao} dense $a$-C matrices. 

Despite the importance of amorphous and nanocomposite carbon systems, 
several critical aspects of their elastic properties and strength remain 
elusive. The elastic moduli of $ta$-C and of ``amorphous diamond'' 
($a$-D), a hypothetical 100\% $sp^3$-bonded model, were calculated 
to be lower than diamond's \cite{KelPRL94}. 
This indicates that continuous random networks are softer than the 
equivalent crystal, despite containing the same bonds, but
the cause is unclear. Even more, the deformation properties 
beyond the elastic regime of $ta$-C, widely used in applications as a 
hard coating, and of the equivalent nanocomposite $n$-D/$a$-C 
are not yet known.

Some of the fundamental issues requiring investigation are: 
(a) It is well known that diamond's strength is limited by the existence
of easy-slip (cleavage) planes \cite{Telling}. Such planes do not 
exist in $a$-D, or $ta$-C. Would this make the latter yield 
at higher stress than diamond and, if not, what is the cause of
weakening? (b) The elucidation of fracture mechanisms in both $ta$-C 
and $n$-D/$a$-C. Especially, one asks whether fracture in $n$-D/$a$-C 
occurs inter- or intra-grain, and which are the weak links leading 
to fracture. (c) Ultimately, is there any enhancement of elastic moduli 
and strength of the nanocomposite $n$-D/$a$-C compared to the pure
$ta$-C phase? Similar questions can be asked for any 
nanostructured material.

Here, we present the results of state-of-the-art tight-binding molecular 
dynamics (TBMD) simulations, which shed light into these fundamental
issues. We study both the elastic properties and the ideal strength and 
fracture mechanisms in amorphous and nanostructured carbon. 
We find that fracture in the nanocomposites occurs inter-grain, in the 
embedding matrix, and so the ideal strength is similar to that of 
$ta$-C. However, their elastic moduli are significantly enhanced over 
$ta$-C. The onset of fracture takes place at weakly bonded $sp^3$ sites.

The TB method is well suited for the present problem. It is more accurate 
than empirical schemes because it provides a quantum mechanical 
description of the interactions. On the other hand, while less accurate than 
{\it ab initio} approaches, it yields greater statistical precision and 
allows the use of larger cells.

We use the TB method developed at the Naval Research Laboratory 
(NRL) \cite{Cohen94}. 
This is a non-orthogonal model, using distance- and environment-dependent 
parameters for transferability between different structures. 
The $sp$ parameters for C were obtained by fitting to a
first-principles database, which includes both the band structure 
and energies of the diamond, graphite, sc, bcc, and fcc configurations, 
as well as of the C$_2$ dimer. The fitting of the band structure (as a
test, the predicted band structure of C$_{60}$ matches independent LDA 
calculations) provides additional robustness to the TB Hamiltonian. 
This is an important advantage 
of the present TB method, which enforces the transferability of these 
parameters to the amorphous state. The fitting was extended to large 
nearest-neighbor distances, up to 6.6 \AA, so the approach properly 
describes the quantum mechanical effects of bond breaking and microfracture.
The calculated phonon frequencies and elastic constants of diamond are 
reasonably close to experimental values. The equilibrium bulk, 
shear, and Young's moduli and the density of diamond are calculated 
at 480 (443) GPa, 494 (476) GPa, 1300 (1145) GPa, and 
3.65 (3.51)g cm$^{-3}$, respectively, compared to the experimental values 
given in parentheses.
For a review of the NRL/TB scheme, see Ref.\ \cite{Papacon03}.

We use supercells of 512 atoms with periodic boundary conditions
(PBC). The $n$-D/$a$-C structures contain a spherical nanocrystal in the 
middle, surrounded by dense amorphous carbon. It is essential to 
use cells of this size, something not feasible with first-principles MD 
methods, in order to describe the nanoinclusions properly, with respect 
to realistic sizes and volume fractions. Smaller cells are not realistic 
and are inadequate.

The nanocomposite structures were generated 
by melting and subsequent quenching at constant volume, in the canonical
($N,V,T$) ensemble, of a diamond structure, while keeping the
atoms in the central spherical region frozen in their positions \cite{Fyta03}.
After quenching, producing amorphization of the surrounding matrix, the
cells are fully relaxed with respect to atom positions and volume. The
density and coordination of the matrix is controlled by the initial 
starting volume of the supercell. The pure $a$-C cells are also generated
by quenching from the melt. In both cases, the liquid was prepared at 6000 K,
and typical quenching durations and rates used are 40 ps and 300 K/ps, 
respectively. In addition, we examine the properties of the 
Wooten-Winer-Waire (WWW) model \cite{WWW} of $a$-D, which serve as a 
benchmark for the calculated properties of $a$-C and $n$-D/$a$-C networks.
The WWW structure (512 atoms) is fully relaxed with the NRL/TBMD approach.
All calculated properties are inferred at 0 K.

Two of the TBMD-generated structures, at equilibrium, are shown in 
Fig. 1. Panel (a) portrays a typical nanocomposite network.
Due to the PBC, this corresponds to a special case, where we have 
a homogeneous dispersion of crystallites of equal size in the matrix, 
at regularly ordered positions. The diamond nanocrystal has a diameter of 
12.5 \AA \, and its volume fraction is 31\%.
The density of the matrix $\rho_{am}$ is 3 g cm$^{-3}$, and its mean
coordination $\bar{z}_{am}$ is 3.8. Panel (b) shows an equivalent 
single-phase $ta$-C network with the same $sp^3$ content and density. 
The $sp^2$ sites are largely clustered. The network contains a 
considerable fraction of both three- and four-membered (3-$m$ and 4-$m$) 
rings, as computed using the shortest-path criterion of 
Franzblau \cite{Franz}. Specifically, there are 20 3-$m$ and 38 4-$m$ 
rings. The vast majority of sites involved in such small rings are 
$sp^3$, $\sim$ 95\%. The average bond length in 3-$m$ (4-$m$) 
rings is 1.50 (1.55) \AA. These characteristics are in excellent agreement 
with what state-of-the-art {\it ab initio} MD simulations 
predict \cite{Marks-McCull}. For example, in a network of a similar 
density with 125 atoms, three 3-$m$ and eleven 4-$m$ rings, composed of 
$sp^3$ atoms, with average bond lenghts of 1.5 - 1.6 \AA \, were found. 
This comparison shows that the present TB method
treats accurately the strain energy of $sp^3$ sites in small rings.
Also, the density--coordination ($sp^3$ fraction) relation for the 
$a$-C networks is found to be linear, in agreement with experiment and
recent TBMD simulations using a different TB Hamiltonian \cite{Mathiou}.

We begin our analysis by examining the elastic properties of the
networks. As a representative quantity, we calculated their bulk
modulus $B$ as a function of the $sp^3$ fraction in
the cells, by fitting the energy versus volume curves with the 
Murnaghan equation of state. Fig. 2 shows the variation of $B$ for 
various nanocomposite structures and for single-phase $a$-C. 
The values for diamond and $a$-D are shown for comparison. The moduli of 
$ta$-C networks ($sp^3$ fraction $\geq$ 60\% \cite{Fyta03}) are quite high,
and that of $a$-D reaches $\sim$ 90\% of the diamond value, confirming 
previous calculations \cite{KelPRL94,Mathiou}. Clearly, the moduli
of $n$-D/$a$-C networks are considerably higher than those of $a$-C.
A mean increase over 10\% is evident. The enhancement becomes stronger
as the $sp^3$ fraction, or the density, increases. A similar
effect is seen as the size, or volume fraction, of the nanocrystals 
increases. Impressively, beyond a certain point, $B$'s are shown to 
exceed the $a$-D value and approach that of diamond. 
We conclude that the elastic response of the composite structures to
hydrostatic deformation is controlled by the nanoinclusions.

We now turn to the calculations of the ideal strength \cite{Zhang,Blase}
of the materials under study. This is the maximum stress
that a material can sustain under non-hydrostatic loads before becoming 
unstable and yielding to plastic deformation or fracture. In order
to check the validity of the employed method, we calculated the ideal
tensile and shear strengths for diamond. From the stress--strain 
curves (not shown), and in the case of tensile load, we found
209 (225) GPa for the $\langle$100$\rangle$ direction, 130 (130) for the 
$\langle$110$\rangle$, and 124 (95) GPa for the $\langle$111$\rangle$ 
direction. In the case of shearing, we found 130 (93) GPa for the 
\{111\}$\langle$112$\rangle$ slip system. Our results
are in good agreement with the {\em ab initio} results of Telling 
{\em et al.} \cite{Telling}, given in parentheses, and confirm that
the \{111\} plane is the easy-slip plane in diamond. The corresponding 
tensile strain at the maximum strength is 0.28, 0.21 and 0.15 for 
$\langle$100$\rangle$, $\langle$110$\rangle$ and $\langle$111$\rangle$, 
respectively. The critical shear strain is 0.23.

Having established the reliability of the method, we proceed to the
study of our amorphous and nanocomposite structures.
For the latter, we apply tensile load in the $\langle$111$\rangle$ 
direction and shear load on the \{111\} plane in the $\langle$112$\rangle$ 
direction. The crystallographic directions are those of the nanodiamond 
region. In both cases, we choose the easy-slip plane of diamond, so that 
we can have a direct comparison to diamond, and examine the
possibility that such planes play a significant role in 
nanocomposites. The a-C phases are highly isotropic, therefore all 
directions of tensile or shear load are equivalent.

The structures were strained in a series of incremental strains.
At each step, atomic positions were fully relaxed with TBMD. The
stress was extracted by differentiating the energy with respect to
strain. The resulting stress-strain data points and fitted curves,
for both tensile and shear load, are shown in Fig. 3. 
There are a number of important aspects of these results. The
first is that all structures have significantly lower ideal strengths and 
critical strains than diamond, including the $a$-D model which is 
100\% $sp^3$-bonded and has no planes of cleavage. 

Obviously, the difference stems from the weaker character of 
the C--C bond in the amorphous network.
Note that while the modulus of $a$-D reaches 90\% that of
diamond, its strength is only about half. This indicates softer
angular forces, due to dihedral disorder, which render the $sp^3$ 
hybrids more easily unstable under non-hydrostatic stresses at the 
bond-breaking regime, far from equilibrium. On the other hand, 
radial rigidity near equilibrium is not so weakened, giving rise to 
high elastic moduli. In the case of $ta$-C, we have, in addition, 
the presence of $sp^2$ sites, which might contribute to the further 
decrease of strength, as it is evident in Fig. 3. We analyze these issues 
more deeply below.

Our most striking finding is that 
the ideal strength of $n$-D/$a$-C is about equal to that of $ta$-C, when
having equivalent amorphous networks, for both tensile and shear loads. 
It indicates that the nanoinclusions do not actually contribute
to the increase of strength, but that this is rather controlled by the other
part of the composite structure, the amorphous matrix.

In order to understand this result, we examined the fracture mechanisms
in our networks. For this purpose, we analyzed the microstructure
just after the critical strain when bonds start to break. We consider
a bond as being broken when its length has become longer than the
first minimum in the pair distribution function of the network at
equilibrium \cite{Galli99}. 
The so extracted cutoff distances are 1.90 \AA, for the amorphous networks, 
and 1.75 \AA \, for the nanodiamond crystallite. (For $n$-D/$a$-C, the 
partial functions in the two components were extracted.)

Fig. 4 shows ball and stick models of $ta$-C and nanocomposite carbon 
under critical tensile strain. Atoms that have lost at least one 
bond are distinguished, according to their hybridization, from atoms
with no broken bonds. It is clear that bond breaking in the
nanocomposite takes place in the amorphous matrix and not in the
nanodiamond, i.e., fracture occurs inter-grain and not intra-grain.
Only at the interface some originally elongated nanodiamond bonds are
broken. (Stress in the nanodiamonds is mostly concentrated towards
the interface \cite{Fyta03}.)
This picture explains why the ideal strengths and corresponding 
strains of $n$-D/$a$-C and $ta$-C are about equal. The fracture
mechanisms in both materials are similar.

Interestingly, the vast majority of atoms involving broken bonds are 
$sp^3$ hybrids, in both the $n$-D/$a$-C and $ta$-C materials. 
This can be understood on the basis of the relative energetics of 
$sp^3$ and $sp^2$ hybrids in the amorphous phase. It has been
previously shown \cite{KelPRL92} that the average atomic energies of 
$sp^3$ and $sp^2$ sites in $ta$-C are -6.8 and 
-7.1 eV, respectively, with respect to atomic C. This is a huge energy 
difference between the two hybrid states compared to the 
almost degenerate energies in diamond and 
graphite ($\sim$ -7.4 eV). Therefore, in the presence of critical strain 
at the onset of fracture, when bonds have to break, it will be 
energetically favorable to loose bonds belonging to $sp^3$ atoms in the 
amorphous region. It is clear from Fig. 4 that microfracture originates
in $sp^3$-rich clusters.

Then, what is the role of $sp^2$ sites, and why the strength of $ta$-C
is lower than that of the all $sp^3$-bonded $a$-D? A close inspection
of the $ta$-C network in Fig. 4(a) shows that the reduction is not directly
linked to $sp^2$ sites, as only few of them have broken bonds, but 
indirectly through their spatial association with $sp^3$ sites. The
fracture region involves many clustered $sp^2$ sites, but the weak links
are exactly those $sp^3$ sites in their immediate vicinity. This is
because these $sp^3$ atoms are under mostly tensile stress \cite{KelPRL94},
contrary to the majority of $sp^3$ sites which are under the more favorable 
compressive conditions, and so they have higher atomic 
energies \cite{KelPRL92}. The unstable sites under strain easily
give up a bond to become $sp^2$. This is the favored hybrid state on the 
surfaces \cite{KelJNCS} ultimately formed by fracture, when the 
material breaks apart. The absence of $sp^2$ sites in the WWW
model, acting as precursors for the destabilization of $sp^3$ sites,
explains its higher strength compared to $ta$-C. That the $sp^2$ sites
are successible to larger plastic deformations is evident in the case 
of the 50\%-50\% $a$-C cell, as shown in Fig. 3.

Finally, by extracting the yield stress $Y$ (defined as the stress where the
strain departs 0.2\% from linearity) and the Young's moduli $E$ from the
stress-strain curves of Fig. 3, and by using the empirical relation of
Tabor \cite{Tabor}, $H/Y = 0.07 + 0.6\, ln(E/Y)$, we estimated the hardness 
$H$ of our networks. We find $\sim$ 90 GPa for $a$-D (WWW), $\sim$ 70 GPa for
$ta$-C and $n$-D/$a$-C, and $\sim$ 40 GPa for 50\%-50\% $a$-C,
compared to 120 GPa for diamond with the present TB method. The value
for $ta$-C is in accord with the experimentally reported values. 
The nanocomposites offer a clear advantage over $ta$-C, namely
their superior elastic properties. They are also expected to 
have excellent thermal stability, similar to that shown by $ta$-C, 
and relaxed intrinsic stress, as proposed elsewhere\cite{Fyta03}. 
Experimentally, stress relief can be achieved by moderate thermal 
annealing, as for $ta$-C. The combination of these properties make 
nanocomposite carbon materials very useful for MEMS/NEMS devices. 

We are grateful to Shay Lifshitz for providing us with a preprint of
the paper in Ref.\ \cite{Yao}.
This work is supported by a grant from the EU and the Ministry of National
Education and Religious Affairs of Greece through the action
``E$\Pi$EAEK'' (programme ``$\Pi\Upsilon\Theta$A$\Gamma$OPA$\Sigma$''.)

\newpage

\begin{center}

\begin{figure}
\includegraphics[width=0.8\textwidth]{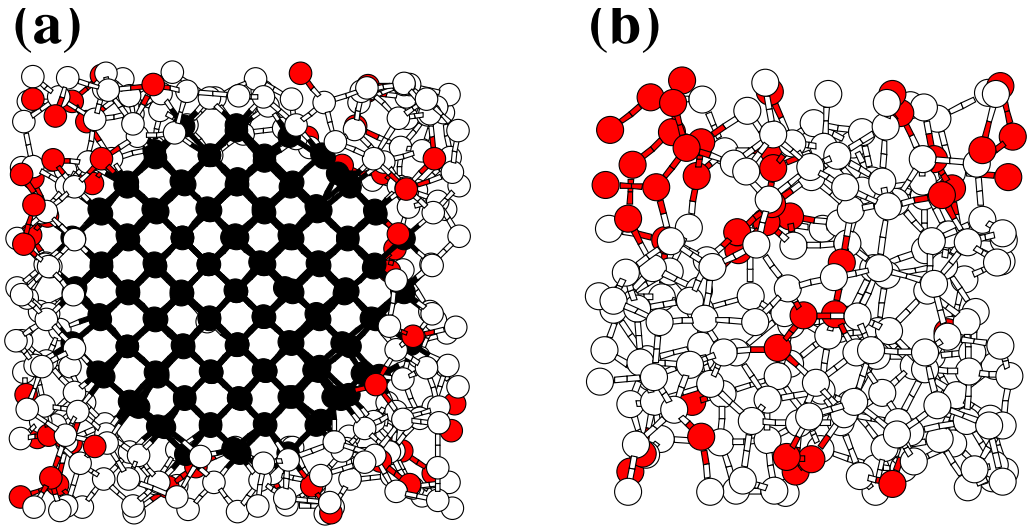}
\caption{(Color online) Ball-and-stick models for (a) a $n$-D/$a$-C network, 
and (b) a pure $ta$-C structure. Red (empty) spheres denote $sp^2$
($sp^3$) sites. In (a), atoms belonging to the nanocrystal are
represented by black spheres.}
\end{figure}


\begin{figure}
\vspace*{1cm}
\includegraphics[width=0.5\textwidth]{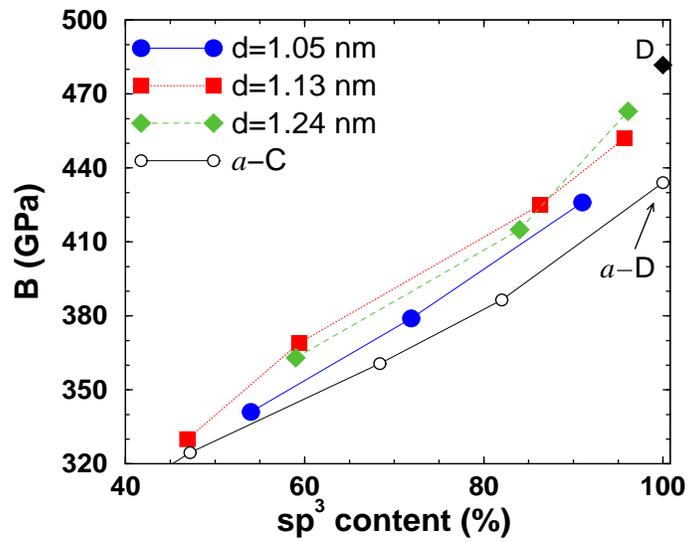}
\caption{(Color online) Bulk modulus as a function of $sp^3$ content for 
$a$-C and $n$-D/$a$-C, with nanodiamonds of different diameter $d$.
Calculations for bulk diamond (D) and $a$-D (the fully tetrahedral 
WWW model of a-C) are also shown for comparison.}
\end{figure} 



\begin{figure}
\includegraphics[width=0.9\textwidth]{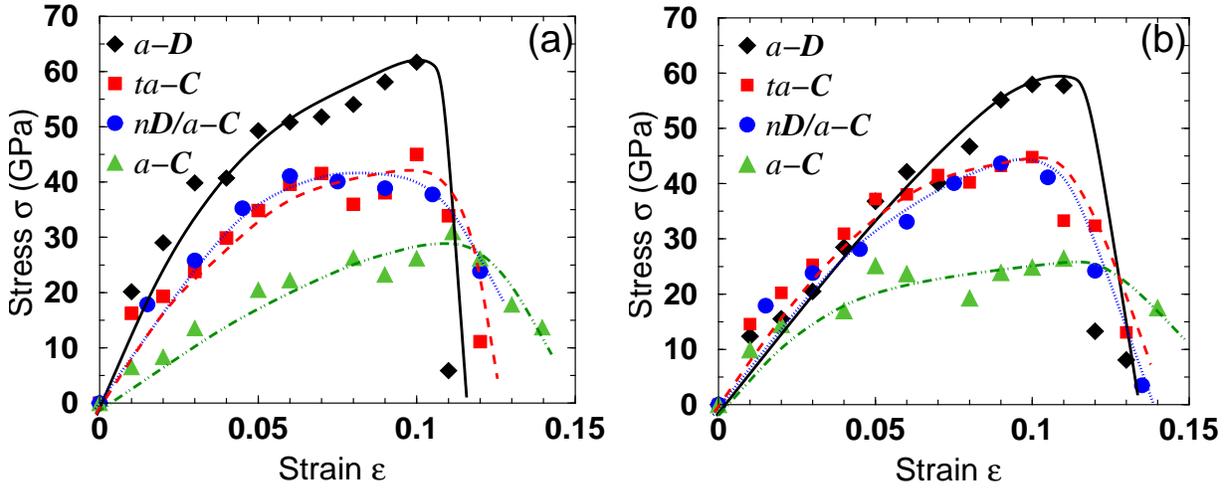}
\caption{(Color online) Stress versus strain curves for the various 
structures studied in this work. The $ta$-C and $n$-D/$a$-C contain 80\%
$sp^3$ sites in the amorphous matrix; the $a$-C cell contains
50\% $sp^3$ sites. Lines are fits to the data points.
(a) Tensile load in the (111) direction. (b) Shear load on the [111] 
plane in the (112) direction.}
\end{figure}
 


\begin{figure}
\vspace*{1cm}
\includegraphics[width=0.9\textwidth]{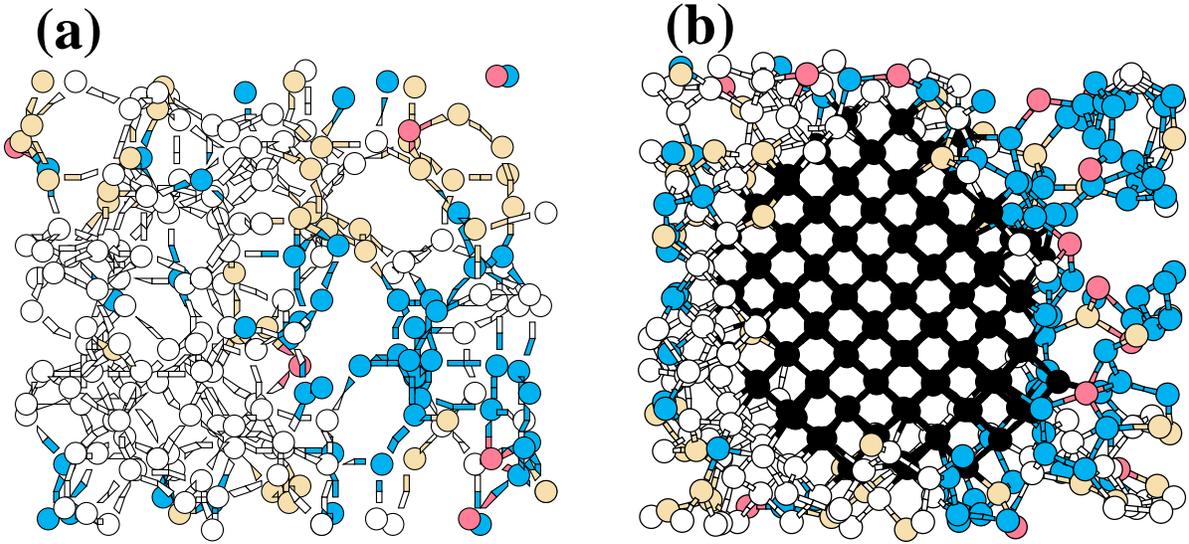}
\caption{(Color online) Ball-and-stick models for fracture in 
(a) single-phase $ta$-C and (b) nanocomposite $n$-D/$a$-C, of similar 
densities. Beige and empty spheres denote $sp^2$ and $sp^3$ 
sites with no broken bonds, respectively. Dark spheres show atoms of 
the nanodiamond. Pink and blue spheres denote broken $sp^2$ and $sp^3$ 
sites, which lost at least one bond, respectively.}
\end{figure} 
\end{center}
\end{document}